\documentclass[twocolumn,showpacs,floats,floatfix,superscriptaddress,aps,pra]{revtex4-1}
\usepackage{amsfonts}
\usepackage{amssymb}
\usepackage{amsmath}
\usepackage{graphicx}
\usepackage{bm}
\usepackage{color}
\usepackage{epsfig}
\usepackage{calc}
\usepackage{ifthen}
\usepackage{subfigure}
\usepackage{graphicx}
\usepackage{epstopdf}
\usepackage{epsfig}

\begin{document}

\title{Long distance adiabatic  wireless energy transfer via multiple coils coupling}
\date{\today }

\begin{abstract}
Recently, the wireless energy transfer model can be described as the Schrodinger equation [Annals of Physics, 2011, 326(3): 626-633; Annals of Physics, 2012, 327(9): 2245-2250]. Therefore, wireless energy transfer can be designed by coherent quantum control techniques, which can achieve efficient and robust energy transfer from transmitter to receiver device. In this paper, we propose a novel design of long distance wireless energy transfer via multiple states triangle crossing pattern, which obtains the longer distance, efficient and robust schematic of power transfer. 

\end{abstract}

\pacs{}
\author{Wei Huang}
\affiliation{Guangxi Key Laboratory of Optoelectronic Information Processing, Guilin University of Electronic Technology, Guilin 541004, China}

\author{Xiaowei Qu}
\affiliation{Guangxi Key Laboratory of Optoelectronic Information Processing, Guilin University of Electronic Technology, Guilin 541004, China}

\author{Shan Yin}
\affiliation{Guangxi Key Laboratory of Optoelectronic Information Processing, Guilin University of Electronic Technology, Guilin 541004, China}

\author{Muhammad Zubair}
\affiliation{Department of Electrical Engineering, Information Technology University (ITU) of the Punjab, Software Technology Park, Ferozepur Road, Lahore 54600, Pakistan.}

\author{Chu Guo}
\affiliation{Henan Key Laboratory of Quantum Information and Cryptography, Zhengzhou, Henan 450000, China}

\author{Xianming Xiong}
\email{xlaser$\_$1999@yahoo.com}
\affiliation{Guangxi Key Laboratory of Optoelectronic Information Processing, Guilin University of Electronic Technology, Guilin 541004, China}

\author{Wentao Zhang}
\email{zhangwentao@guet.edu.cn}
\affiliation{Guangxi Key Laboratory of Optoelectronic Information Processing, Guilin University of Electronic Technology, Guilin 541004, China}

\maketitle

%42.82.Et Waveguides, couplers, and arrays
%42.81.Qb Fiber waveguides, couplers, and arrays
%42.79.Gn Optical waveguides and couplers
%32.80.Xx Level crossing and optical pumping

%**************************************************************

\section{Introduction}
The research of wireless energy transfer started at the age of the Tesla. The principle of wireless energy transfer can provide power transfer from one device (transmitter) to another device (receiver) via electromagnetic field and the major advantage of this technique is non-connected power transfer without the wires or cables \cite{Brown1996}. Due to this great dominance, wireless energy transfer technique possesses significant applications in the charging of electric vehicles \cite{Imura2009, Choi2014}, mobile phones \cite{Li2015}, lighting \cite{Jiang2017}, implantable medical devices \cite{Mazzilli2010, Kim2015} and others power supply devices \cite{Barman2015, Sample2011}.

The normal setup of wireless energy transfer is based on two resonant coils with the constraint of exact resonance frequency between the transmitter coil and receiver coil \cite{Hui2013,Zhang2013,Hamam2009,Kurs2007}. Thanks to the coupled mode theory (CMT) of wireless energy transfer \cite{Kurs2007}, the coupling equation of coils can be approximated to Schrodinger equation. Therefore, coherent quantum control technique can be employed to control power transfer between the coils. Recently, two coils coupling \cite{Rangelov2011} and three coils coupling \cite{Rangelov2012} propose an adiabatic technique to enhance the transfer efficiency, comparing with exact resonance method. The adiabatic following is a famous coherent quantum control technique, which is widely used in quantum system and classical systems, such that the quantum state controlling \cite{Huang2017}, quantum manybody processing \cite{Huang2019}, optical waveguide coupler \cite{Alrifai2019}, graphene surface plasmon polaritons (SPPs) coupler and terahertz SPPs coupler \cite{Huang2018,Huang20192}. 

\begin{figure} [htbp]
\centering
\includegraphics[width=0.5\textwidth]{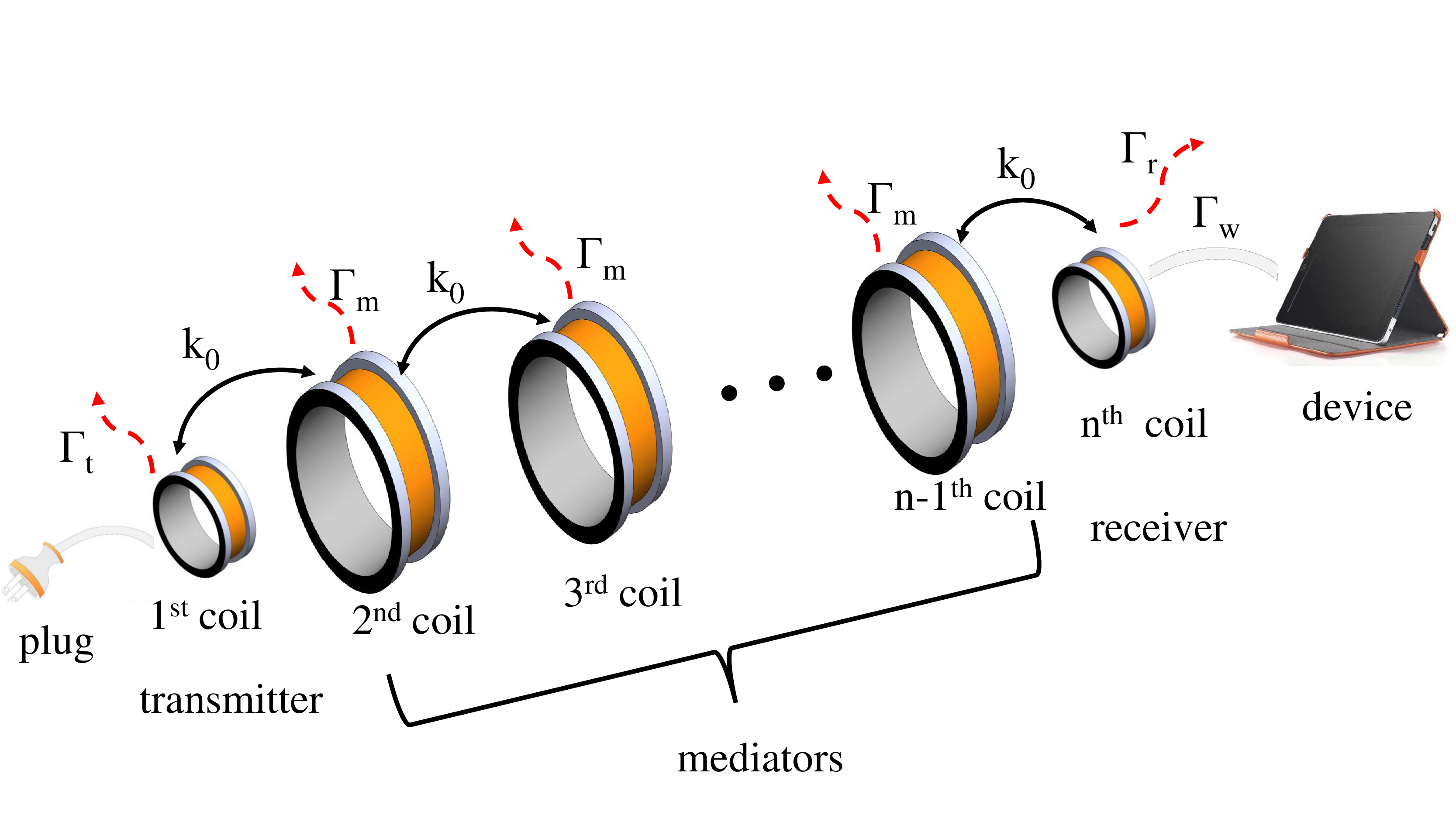}
\caption{The schematic configuration of our designed multiple coils system. The coupling strength between two adjacent coils is $k_0$ and the intrinsic frequencies of transmitter, mediators and receiver are $\omega_{t}$, $\omega_{m}$ and $\omega_{r}$ respectivelly. $\Gamma_{t}$, $\Gamma_{m}$ and $\Gamma_{r}$ are the corresponding intrinsic loss rates, due to absorption and radiation of coils, while $\Gamma_{w}$ is the extraction of work from the receiver.} \label{fig1}
\end{figure}

The coupling strength between two coils decreases exponentially with increasing distance, based on CMT of the coupling between coils \cite{Kurs2007}. Therefore, the limitation of previous researches is that the transmission distance is relatively short due to the two or three coils coupling. In this paper, we propose a novel design of adiabatic wireless energy transfer via multi-coil system. In our design, we set up the multiple mediator coils between transmitter and receiver coil and all the mediator coils are identical. The two adjacent coils are equally spaced which provides equal coupling strength between two adjacent coils. The scheme of our designed configuration is illustrated in Fig. ~\ref{fig1}. In this configuration, we have two significant advancements comparing with previous researches, \textit{(i)} we enhance the power transfer distance by contrasting with others adiabatic following based on two or three coils' coupling due to multiple mediator coils; \textit{(ii)} our design improves the transfer efficiency comparing with exact resonance of multi-coil system. 

\begin{figure} [htbp]
\centering
\includegraphics[width=0.5\textwidth]{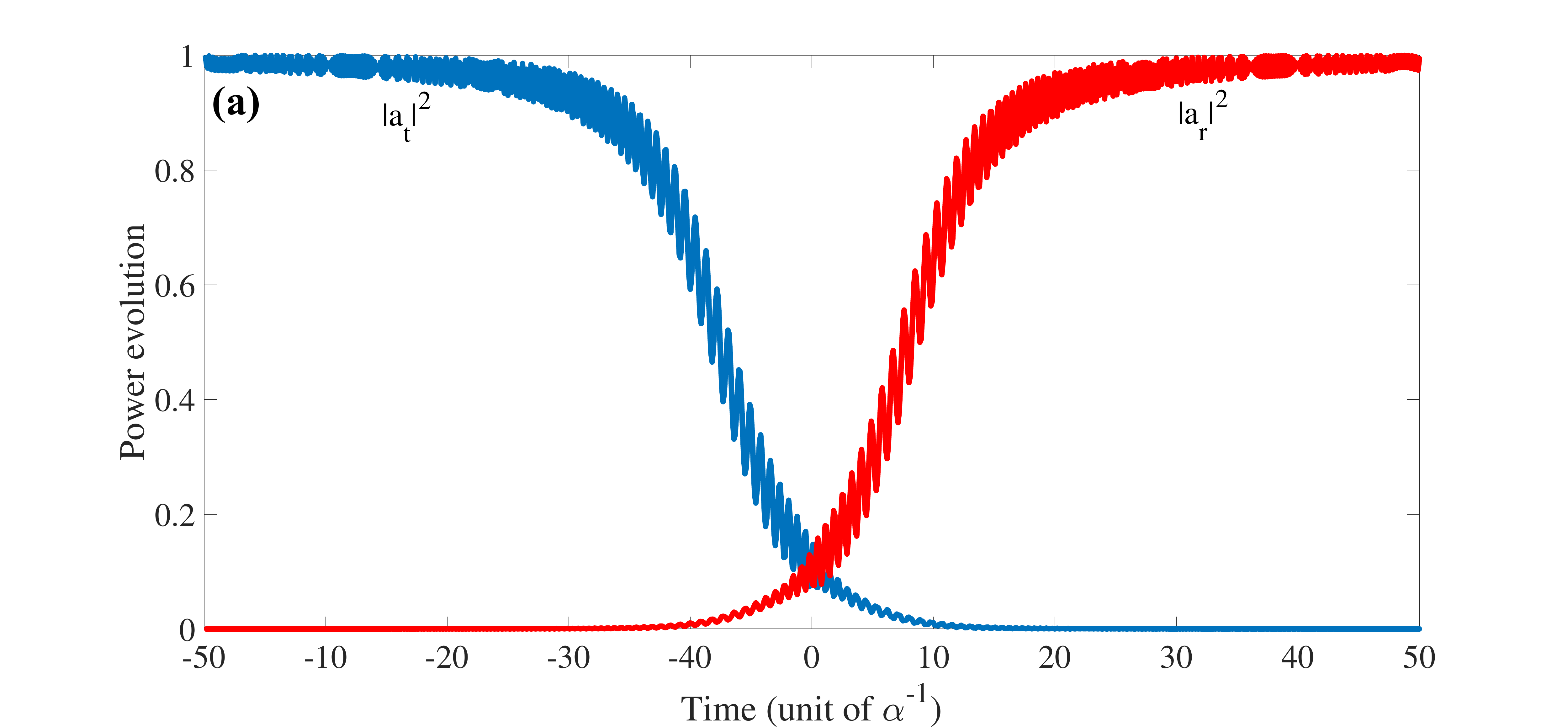}
\includegraphics[width=0.5\textwidth]{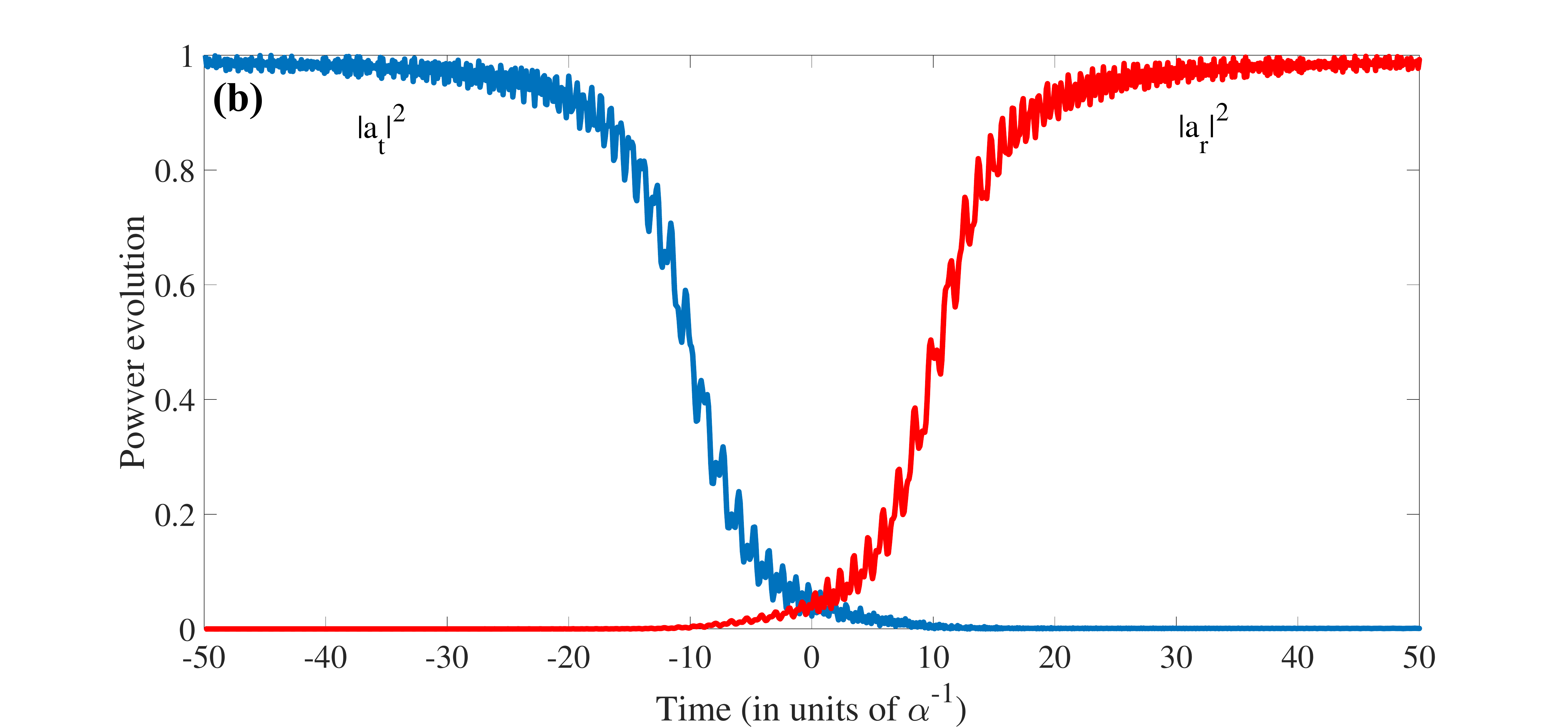}
\includegraphics[width=0.5\textwidth]{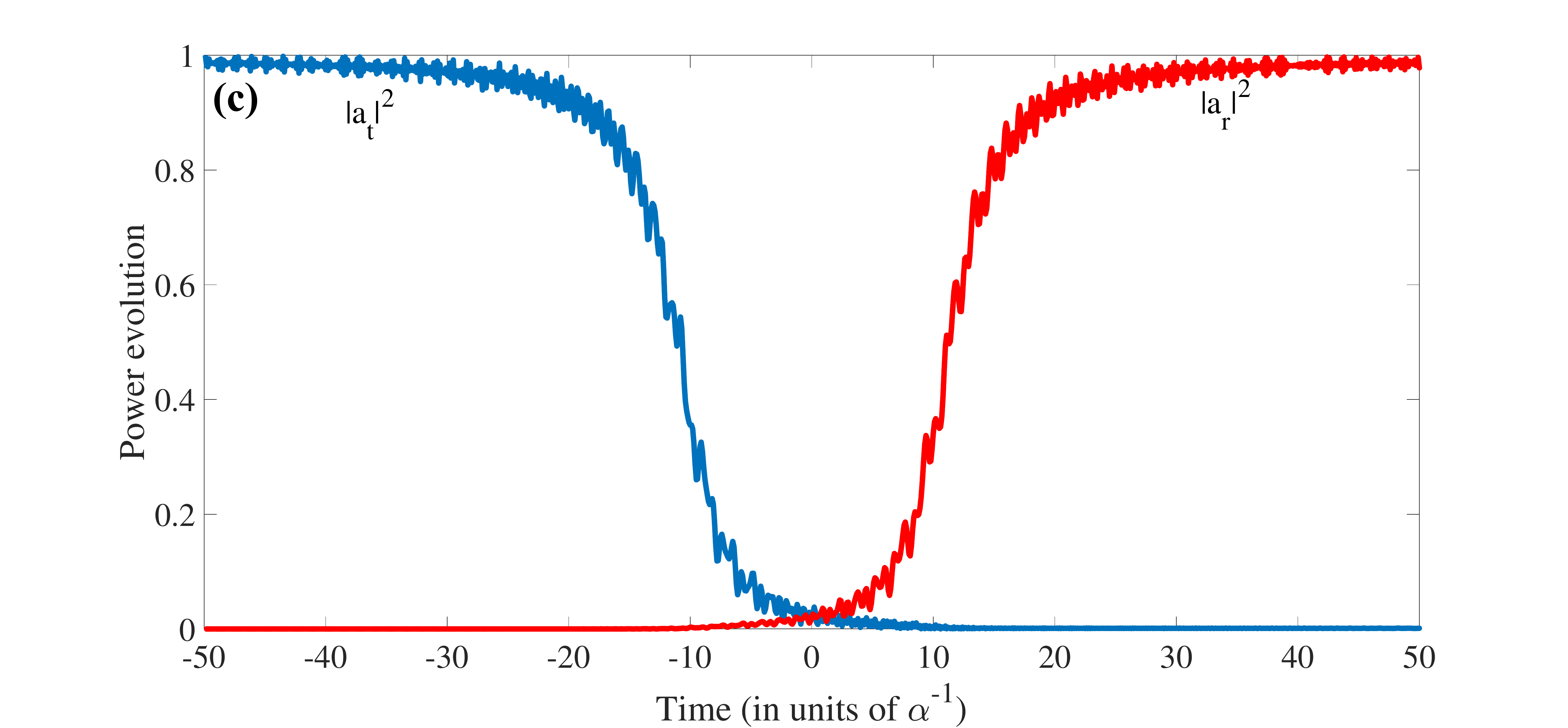}
\caption{The power evolution of transmitter (blue line) and receiver (red line) coil using coupled equation (Eq. 1) without any loss ($\Gamma_t = \Gamma_r = \Gamma_m = \Gamma_w = 0$), with (a) three coils system, (b) four coils system and (c) five coils system. The parameters are given by $\delta = -7 \alpha$, $k_0 = 3.5 \alpha$ and time is from $-50 \alpha^{-1}$ to $50 \alpha^{-1}$.} \label{fig2}
\end{figure}

\begin{figure} [htbp]
\centering
\includegraphics[width=0.5\textwidth]{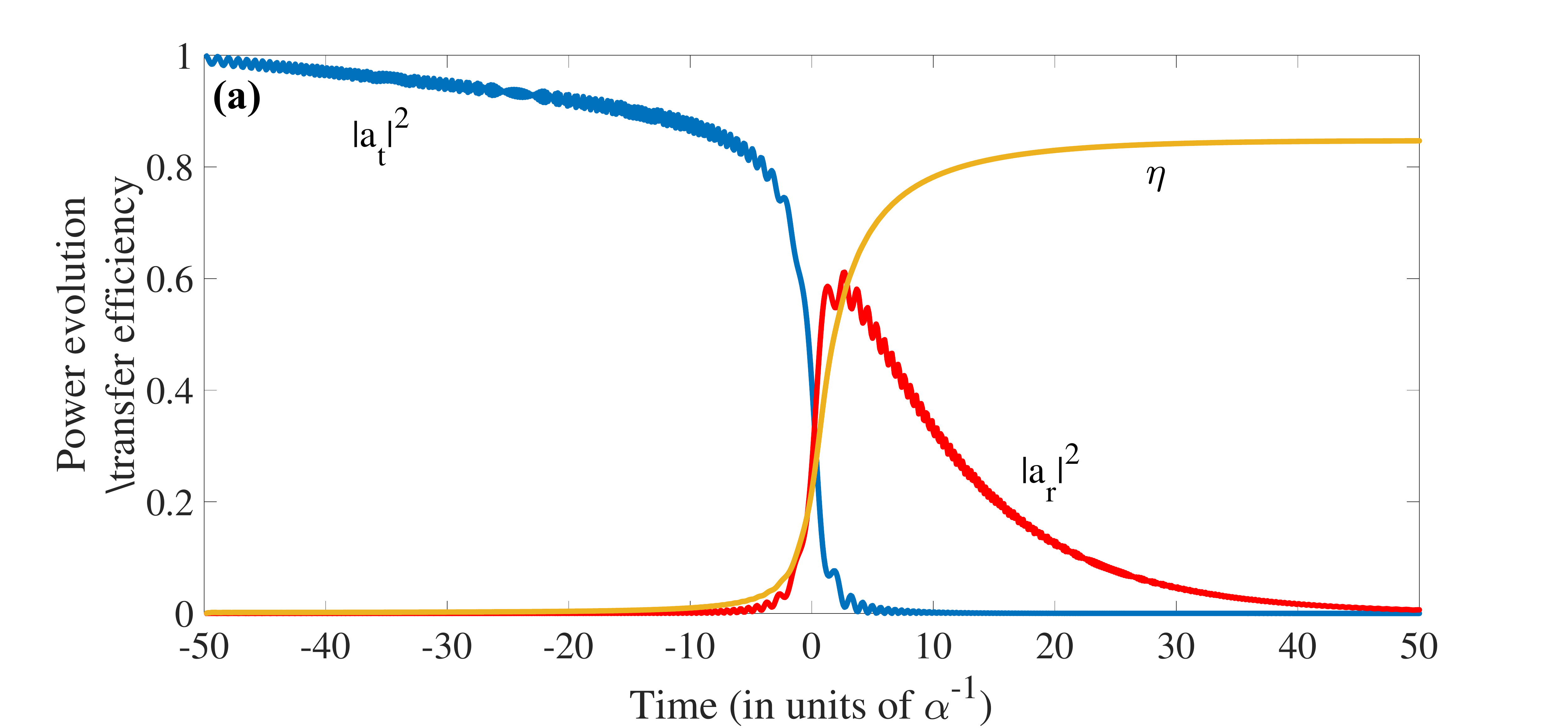}
\includegraphics[width=0.5\textwidth]{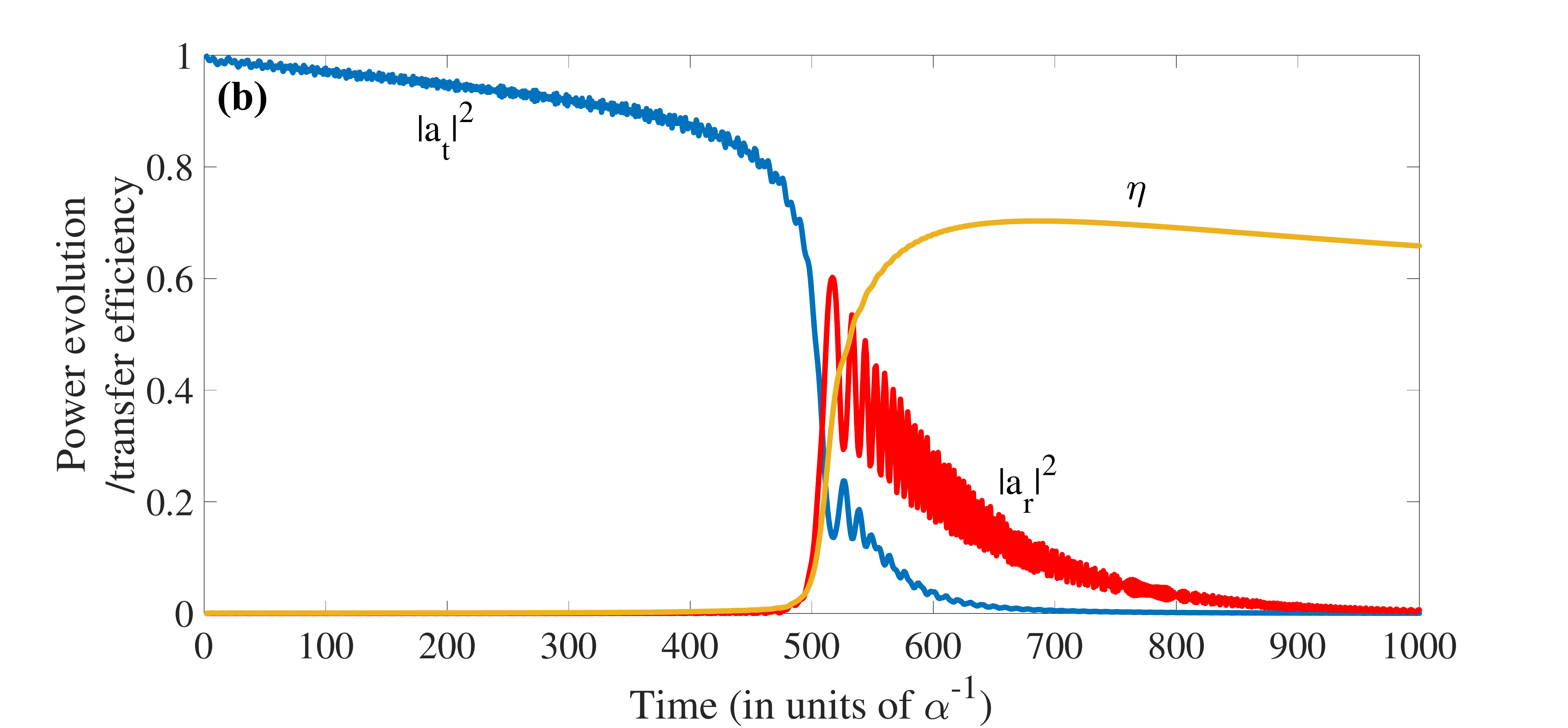}
\includegraphics[width=0.5\textwidth]{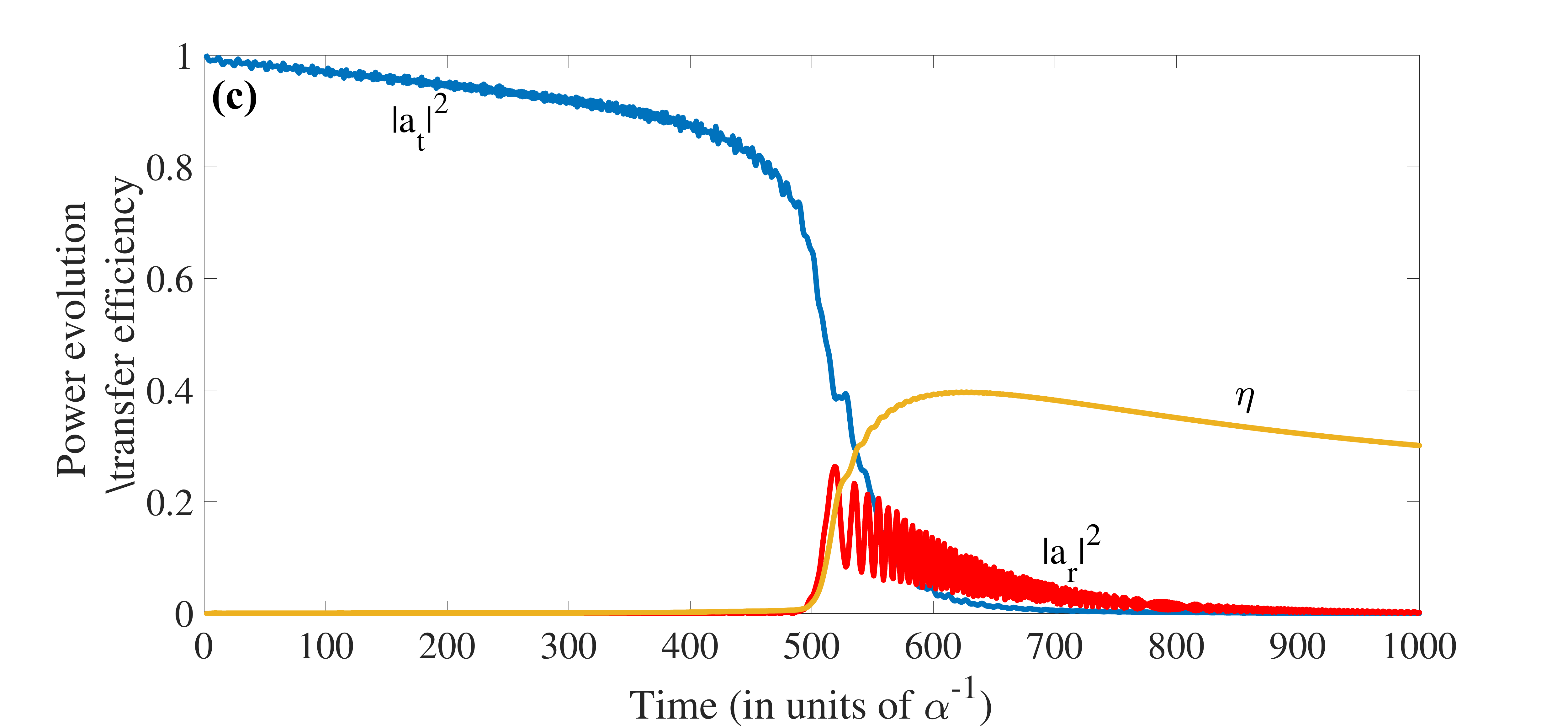}
\caption{The power evolution of transmitter (blue line) and receiver (red line) coil with loss of transmitter and receiver coil $\Gamma_t = \Gamma_r = 0.001 \alpha$, intrinsic loss of mediator coils $\Gamma_m = 0.01 \alpha $ and extracted work rate $\Gamma_w = 0.05 \alpha$, with (a) three coils system, (b) four coils system and (c) five coils system. The transfer efficiency $\eta$ is given by orange line. The parameters are given by $\delta = -7 \alpha$, $k_0 = 3.5 \alpha$ and time is from $-50 \alpha^{-1}$ to $50 \alpha^{-1}$.} \label{fig3}
\end{figure}

In this paper, we design a novel long distance, efficient and robust wireless energy transfer scheme via multi-state triangle crossing pattern, which provides complete power transfer from transmitter coil to receiver coil via multiple mediator coils. We demonstrate that the complete power transfer can be achieved independent of number of coils, without considering any loss in any coil (see Fig. ~\ref{fig2}). Subsequently, we consider the lossy case (the absorption and radiation in all coils and extracted rate from receiver coil) with different number of coils and numerically calculate the transfer efficiency $\eta$ (as shown in Fig. ~\ref{fig3}). Finally, we plot the transfer efficiency $\eta$ against number of coils and different lossy parameters (lossy in the mediator coils and extracted rate), as shown in Fig. ~\ref{fig4}.

\section{Model}
Based on the CMT of coupling between coils, we can easily express the coupled equation of muti-coil into multi-state Schrodinger equation, written as 
\begin{equation}
i\dfrac{d}{d t}
\begin{bmatrix}
a_{t} (t) \\
a_{2} (t) \\
\vdots \\
a_{n-1} (t) \\
a_{r} (t)
\end{bmatrix}
= \textbf{H}(t) \begin{bmatrix}
a_{t} (t) \\
a_{2} (t) \\
\vdots \\
a_{n-1} (t) \\
a_{r} (t)
\end{bmatrix}.
\end{equation}
where $a_{n}$ is the power amplitude on the $n^{th}$ coil, with power $P_{n} = |a_n|^2$ and \textbf{H} is the Hamiltonian of the multi-coil coupling system, which is, 
\begin{equation} 
\textbf{H} =\begin{bmatrix}
\omega_{t} - i \Gamma_{t} & k_0 & \ddots \\
k_0 & \omega_{m} - i\Gamma_{m} & k_0 \\
 & \ddots & \ddots & \ddots & \\
 & & k_0 & \omega_{m} - i\Gamma_{m} & k_0 \\
 & & & k_0 & \omega_{r} - i \Gamma_r - i \Gamma_w \\
\end{bmatrix},
\end{equation}
where $\omega_{t}$, $\omega_{m}$ and $\omega_{r}$ are the intrinsic frequencies of transmitter, mediators and receiver, with $\omega_{i} =1 /\sqrt{L_i(t)C_i(t)}$. $L_i(t)$ and $C_i(t)$ are the inductance and the capacitance of $i^{th}$ coil. $\Gamma_{t}$, $\Gamma_{m}$ and $\Gamma_{r}$ are the corresponding intrinsic loss rates, due to absorption and radiation of coils, while $\Gamma_{w}$ is the extraction of work from the receiver. In addition, the coupling strength between $(i-1)^{th}$ and $i^{th}$ coil can be given by CMT, written as $k_0 = M \sqrt{(\omega_{i-1}\omega_{i})/(L_{i-1}L_{i})}$, where $M$ is the mutual inductance of the two coils. In our design (the scheme as shown in Fig.~\ref{fig1}), we choose the variable inductance $L$ and capacitance $C$ to vary the intrinsic frequencies of transmitter $\omega_{t}$ and receiver $\omega_{r}$ via external control and the intrinsic frequencies of all mediator coils are constant $\omega_m$. We set up the transmitter coil, mediator coils and receiver coil with fixed distance. Therefore, the coupling strength between adjacent two coils is constant (time independent), $k_0$. 

Subsequently, it is well known that the definition of transfer efficiency $\eta$ is the ratio between the extract work from the receiver coils divided by the total energy loss off the system \cite{Rangelov2011,Rangelov2012}, given by
\begin{equation}
\eta = \dfrac{\Gamma_w \int^{T}|a_r|^2 dt}{\Gamma_t \int^{T}|a_t|^2 + \sum \Gamma_m \int^{T}|a_m|^2 + (\Gamma_r+\Gamma_w) \int^{T}|a_r|^2 dt}.
\end{equation}

The previous researches had already shown the complete quantum transition between three quantum levels via triangle crossing pattern \cite{Unanyan2001, Ivanov2008}, which forces the quantum evolution along with one adiabatic state of the system. Their Hamiltonian is special case (three-level state) of our designed Hamiltonian $\textbf{H}$ (see Eq. 2) and our Hamiltonian $\textbf{H}$ is the multi-state chains of N quantum state, which has only coupling between its two neighbors of each state, such that $1 \leftrightarrow 2 \leftrightarrow ... \leftrightarrow N-1 \leftrightarrow N$. In this configuration, we can transfer our multi-level quantum system to three-level state like with a dressed middle state \cite{Vitanov2001}. Therefore, we can produce complete transfer population to our designed multiple quantum states system $\textbf{H}$ (as shown in Eq. 2) with triangle crossing pattern, while the frequency of the transmitter and receiver coil change in time in opposite direction and frequencies of mediators are the constant, given by \cite{Rangelov2012} 
\begin{equation}
\begin{aligned}
\omega_t &= \omega_m + \delta - \alpha^2 t ;  \\ 
\omega_m &= \text{constant};  \\
\omega_r &=  \omega_m + \delta + \alpha^2 t.
\end{aligned}
\end{equation}
Therefore, we can employ this frequency shift configuration to produce the complete power transfer from transmitter to receiver in the adiabatic following setting up. There are three transfer patterns within the triangle crossing pattern, such as sequential transfer ($\delta > 0$), bow-tie transfer ($\delta = 0$) and 
direct transfer ($\delta < 0$). The previous researches has already shown that direct transfer pattern has better transfer efficiency $\eta$ \cite{Rangelov2012}. Thus, we will use direct transfer of triangle crossing pattern with $\delta < 0$ in our paper. 

\section{Long distance power transfer}

Subsequently, we can numerically calculate the power evolution of multiple coils coupling system with running coupled equation (Eq. 1) to demonstrate the power transfer from transmitter to receiver and illustrate the transfer efficiency $\eta$ of our design. At the beginning, we demonstrate the result of power evolution of transmitter and receiver without any loss ($\Gamma_t = \Gamma_r = \Gamma_m = \Gamma_w = 0$), in order to illustrate complete power transfer with our design in the multiple coils system. We demonstrate this feature with the parameters $\delta = -7 \alpha$, $k_0 = 3.5 \alpha$ and numbers of coils are three, four, five respectively, as shown in Fig. ~\ref{fig2}. From the results, we easily obtain that the complete power transfer from the transmitter coil to receive coil via the mediator coils, no matter how many mediator coils are. This feature determines the transfer efficiency $\eta$ of our design is larger than exact resonant case, because the energy in receiver coils is transfer back to transmitter coil (so-call Rabi oscillation) in the exact resonant case. 

\begin{figure} [htbp]
\centering
\includegraphics[width=0.5\textwidth]{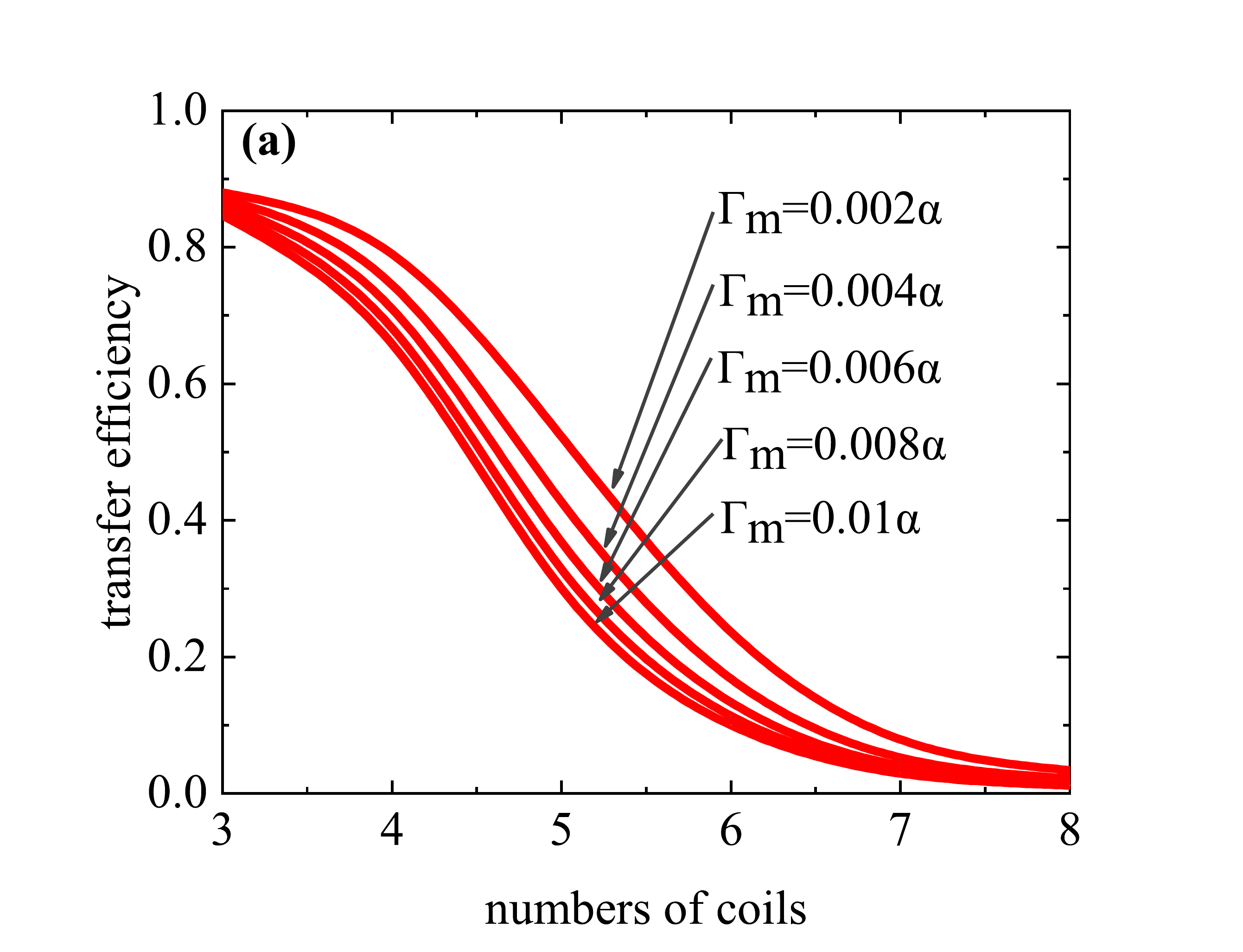}
\includegraphics[width=0.5\textwidth]{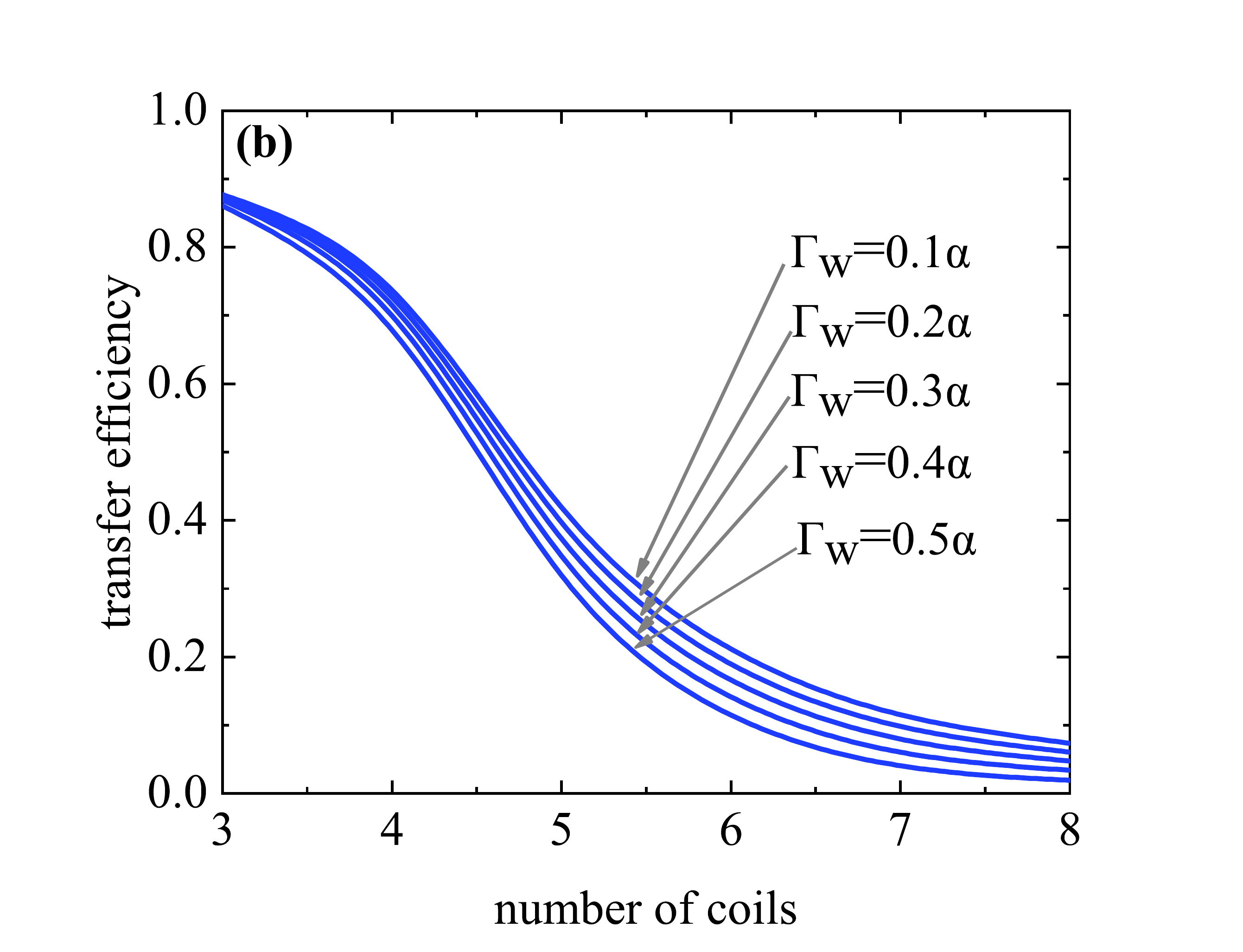}
\caption{The transfer efficiency $\eta$ along with number of coils (a) with different loss of mediator coils $\Gamma_m$ and fixed extracted work rate $\Gamma_w = 0.05 \alpha$; (b) with extracted work rate $\Gamma_w$ and fixed loss of mediator coils $\Gamma_m = 0.01 \alpha$ } \label{fig4}
\end{figure} 

From the Fig. ~\ref{fig2}, we have already demonstrated that our design works functionally in principle. Afterwards, we can numerically calculate the loss case with the parameters intrinsic loss (mainly the absorption) of transmitter and receiver coil $\Gamma_t = \Gamma_r = 0.001 \alpha$, intrinsic loss (mainly the radiation) of mediator coils $\Gamma_m = 0.01 \alpha $ and extracted work rate $\Gamma_w = 0.05 \alpha$, which is suitable for real scenario. Therefore, we obtain the numerical results of power evolution $|a_t|^2$ and $|a_r|^2$ with the loss and plot the transfer efficiency $\eta$ along with the time from $-50 \alpha^{-1}$ to $50 \alpha^{-1}$, as shown in Fig.~\ref{fig3} (a) three-coil (b) four-coil (c) five-coil system. As we can see that the power evolution of transmitter $|a_t|^2$ drops directly along with the time, due to the power transfer continuously flowing from transmitter to receiver. The power on receiver coil starts to increase, and then declines exponentially due to extract work from the receiver. In addition, when the number of coil is increasing, the transfer efficiency $\eta$ (see orange line) decreases, due to the largely loss on mediator coils. 

Furthermore, we plot the transfer efficiency $\eta$ along the number of coils with different loss of mediator coils $\Gamma_m$ and extract work rate $\Gamma_w$. The transfer efficiency largely depends on number of coils, $\Gamma_m$ and $\Gamma_w$, while increasing the number of coils leads to decrease of transfer efficiency $\eta$. When we have multiple mediator coils, the power on the coils flows within the mediator coils. However, the energy on the mediator coils has dissipation. Therefore, it is easily to obtain that improving the quality of mediator coils (increasing the $\Gamma_m$ with fixed $\Gamma_w = 0.05 \alpha$) and increasing the extract work rate $\Gamma_w$ (with fixed $\Gamma_m = 0.01 \alpha$) enhances the transfer efficiency, as shown in Fig. ~\ref{fig4}. 

\section{Conclusion}
In this paper, we propose a novel wireless energy transfer scheme to provide long distance, efficient, robust power transfer via multiple mediator coils, based on multi-state triangle crossing pattern quantum coherent control. Based on numerically calculations, we can illustrate that our design can provide much longer transfer distance (for example transfer distance can increase up to 2 times) with relatively smaller decrease in the transfer efficiency $\eta$ (transfer efficiency $\eta$ drops from $87\%$ to $53\%$). 

\section*{Acknowledgements}
This research was funded by the National Science and Technology Major Project of China (No. 2017ZX02101007-003); the National Natural Science Foundation of China (Nos. 61565004, 61665001 and 61965005); the Natural Science Foundation of Guangxi Province (Nos. 2017GXNSFBA198116 and 2018GXNSFAA281163); the Science and Technology Program of Guangxi Province (No. 2018AD19058); GUET postgraduate outstanding dissertation project (No. 18YJPYSS24). Wei Huang acknowledges the funding from Guangxi Oversea 100 Talent Project and Wentao Zhang acknowledges the funding from Guangxi Distinguished Expert Project.

\end{document}